\documentclass[11pt,a4paper]{article}

\usepackage{amsfonts,amsmath,latexsym}

\title{Quantum and Stochastic Branching Programs of Bounded Width} 

\author{ Farid Ablayev\thanks{Dept. of Theoretical Cybernetics, Kazan
    State University.  {\tt ablayev@ksu.ru }; work done in part while
    visiting Institute of Advanced Study and Max-Plnak Institute for
    Mathematics, supported by RFBR grant 03-01-00769} 
\and 
Cristopher Moore\thanks{ Computer Science Department, University of
  New Mexico, Albuquerque and the Santa Fe Institute {\tt
    moore@cs.unm.edu}; supported by NSF grant EIA-0218563} 
\and
Christopher Pollett\thanks{ Dept. of Computer Science San Jose State
  University.  {\tt pollett@cs.sjsu.edu}} }

\date{}

\usepackage{amssymb}
\usepackage{amsfonts,amsmath,latexsym}

\newcommand{\comment}[1]{}
\newcommand{\Accept}{{\rm Accept}}

\newcommand{\EQN}[1]{\begin{eqnarray*}#1\end{eqnarray*}}
\newcommand{\mat}{\left(\! \begin{array}{cc}}
\newcommand{\rix}{\end{array} \!\right)}
\newcommand{\ket}[1]{{|{#1} \rangle}}
\newcommand{\bra}[1]{{\langle {#1}|}}
\newcommand{\braket}[2]{{\langle {#1} \mid {#2} \rangle}}
\newcommand{\ignore}[1]{}

\newcommand{\eps}{\epsilon}
\newcommand{\Z}{{\Bbb Z}}
\newcommand{\R}{{\Bbb R}}

\newcommand{\nc}{\mathrm{NC}}
\newcommand{\NC}{\mathrm{NC}}
\newcommand{\ACC}{\mathrm{ACC}}

\newcommand{\bp}{\mathrm{BP}}

\newcommand{\qbp}{\mathrm{QBP}}

\newcommand{\QBP}{\mathrm{QBP}}
\newcommand{\SBP}{\mathrm{SBP}}
\newcommand{\BQBP}{\mathrm{BQBP}}
\newcommand{\EQBP}{\mathrm{EQBP}}
\newcommand{\BSBP}{\mathrm{BSBP}}
\newcommand{\BWBP}{\mathrm{BWBP}}

\newcommand{\OBDD}{\mathrm{OBDD}}

\newcommand{\AND}{\mathit{AND}}
\newcommand{\MULT}{\mathit{MULT}}
\newcommand{\MOD}{\mathit{MOD}}

\newtheorem{mydef}{Definition}
\newcommand{\bedef}{\begin{mydef}}
\newcommand{\eedef}{\end{mydef}}

\newtheorem{theorem}{Theorem}
\newcommand{\bethm}{\begin{theorem}}
\newcommand{\eethm}{\end{theorem}}

\newtheorem{conjecture}{Conjecture}

\newtheorem{cors}{Corollary}
\newcommand{\becor}{\begin{cors}}
\newcommand{\eecor}{\end{cors}}

\newtheorem{lemm}{Lemma}
\newcommand{\belem}{\begin{lemm}}
\newcommand{\eelem}{\end{lemm}}

\newtheorem{rft}{Theorem}[section]
\newcommand{\bethmsec}{\begin{rft}}
\newcommand{\eethmsec}{\end{rft}}

\newtheorem{fpr}{Property}
\newcommand{\beprop}{\begin{fpr}}
\newcommand{\eeprop}{\end{fpr}}

\newcommand{\qbps}{\mathrm{QBPs}}
\newcommand{\bps}{\mathrm{BPs}}

\newcommand{\qs}[1]{\mbox{${|\!\!~#1~\!\!\rangle}$}}

\newcommand{\ceiling}[1]{\mbox{$\lceil\!\!~#1~\!\!\rceil$}}

\newcommand{\Endproof}{\hfill$\Box$\\}

\newcommand{\SU}{{\mathit SU}}
\newcommand{\SO}{{\mathit SO}}

\begin{document}

\maketitle
\thispagestyle{empty}

\begin{abstract}

In this paper we show that one qubit polynomial time computations are
 at least as powerful as $\NC^1$ circuits. More precisely,
we define syntactic models  for quantum and stochastic branching
programs of bounded width and prove upper and lower bounds on their
power.  We show any $\NC^1$ language can be accepted exactly by a
width-$2$ quantum branching program of polynomial length, in contrast
to the classical case where width $5$ is necessary unless
$\NC^1=\ACC$.  This separates width-$2$ quantum programs from
width-$2$ doubly stochastic programs as we show the latter cannot
compute the middle bit of multiplication.  Finally, we show that
bounded-width quantum and stochastic programs can be simulated by
classical programs of larger but bounded width, and thus are in
$\NC^1$.

\end{abstract}


\section{Preliminaries}

Interest in quantum computation has steadily increased since Shor's
discovery of a polynomial time quantum algorithm for
factoring~\cite{shor97}.  A number of models of quantum computation
have been considered, including quantum versions of Turing machines,
simple automata, circuits, and decision trees. The goal of much of
this research has been to understand in what ways quantum algorithms
do and do not offer a speed-up over the classical case, and to
understand what classical techniques for proving upper and lower
complexity bounds transfer to the quantum setting.

Branching programs have proven useful in a variety of domains, such as
hardware verification, model checking, and other CAD applications; see
for example the book by Wegener~\cite{we00}.
In addition, branching programs are a convenient model for
nonuniform computation with varying restrictions.
Even oblivious branching programs of constant width --- the 
non-uniform equivalent of finite-state automata --- are surprisingly
powerful.  Indeed, Barrington~\cite{bar85} showed that branching
programs of width 5 are already as powerful as circuits of logarithmic
depth.

Moreover, branching programs are a very natural model for comparing
the power of quantum computation with classical computation, both
deterministic and randomized.  Recently, several models of quantum
branching programs have been proposed
\cite{amp02,agk01,asv00,nahaka00}.

In this paper we define and consider syntactic models of stochastic
and quantum branching programs. For this syntactic model we present
several results for quantum branching programs of bounded width
\cite{amp02}.  We show that width-$2$ quantum programs are more
powerful than width-$2$ doubly stochastic programs, and are as strong
as deterministic branching programs of width $5$.  Specifically, we
show that polynomial-length, width-$2$ quantum branching programs can
recognize any language in $\NC^1$ exactly.  Note that such programs
are equivalent to a nonuniform automaton whose only storage device is
a single qubit!

On the other hand, we show that polynomial-length, width-$2$ doubly
stochastic programs cannot compute the middle bit of the
multiplication function.  In the classical case, Yao~\cite{yao83}
showed that width-$2$ deterministic programs require superpolynomial
length to compute the majority function.

Next, we show that bounded-error (syntactic) quantum and stochastic
programs can be simulated by deterministic programs of the same length
and larger, but still bounded, width. Therefore the class of languages
recognized by these programs coincides with (nonuniform) $\NC^1$.
This also implies that, for bounded-width quantum programs, exact
acceptance is just as strong as acceptance with bounded error.

To give some flavour of what our syntactic model is, consider the usual
computation of a branching program. When we query a variable we do one
of two actions depending on its values. If we query the variable again
we expect to see the same value. An inconsistent state is one that is
reached by querying a variable more than once and using a different
value each time. Our syntactic models allow for the inclusion of
inconsistent final states when the final bounded error acceptance
property is calculated.  This enables us to show that final consistent
states of stochastic and quantum programs recognizing languages with
bounded error recognition have a specific metric property: sets of
accepting and rejecting states are isolated from each other. We define
syntactic programs as programs which satisfy the property that the
set of all (consistent as well as inconsistent) final states form a
partition into two sets isolated from each other.

We call this property a syntactic property in analogy with the
notion of syntactic classical read-$k$-times programs. The definition of
classical syntactic read-$k$-times branching programs \cite{brs93}
includes all their nonconsistent paths in the general read-$k$-times
variables testing property.

We use the techniques of our result to show that polynomial-length
width-$2$ stochastic programs that accept with probability $1/2+\eps$
cannot compute the majority function if $\eps > 1/4$. In addition, we
show that polynomial-length stochastic programs with width $2$ and
$\eps > 1/8$, width $3$ and $\eps > 1/3$, or width $4$ and $\eps >
3/8$ can only recognize languages in $\ACC$.

\section{Branching Programs}

We begin by discussing the classical model of branching programs and
then generalize it to the quantum setting. A good source of
information on branching programs is Wegener's book~\cite{we00}, and
for an introduction to quantum computation see Nielsen and
Chuang~\cite{nc00}.

\bedef
A {\em branching program} is a finite directed acyclic graph which
accepts some subset of $\{0,1\}^n$.  Each node (except for the sink
nodes) is labeled with an integer $1 \le i \le n$ and has two outgoing
arrows labeled $0$ and $1$.  This pair of edges corresponds to
querying the $i$'th bit $x_i$ of the input, and making a transition
along one outgoing edge or the other depending on the value of $x_i$.
There is a single source node corresponding to the start state, and a
subset $\Accept$ of the sink nodes corresponding to accepting states.
An input $x$ is {\em accepted} if and only if it induces a chain of
transitions leading to a node in $\Accept$, and the set of such inputs
is the language accepted by the program.  A branching program is {\em
oblivious} if the nodes can be partitioned into levels $V_1, \ldots,
V_\ell$ and a level $V_{\ell+1}$ such that the nodes in $V_{\ell+1}$
are the sink nodes, nodes in each level $V_j$ with $j \le \ell$ have
outgoing edges only to nodes in the next level $V_{j+1}$, and all
nodes in a given level $V_j$ query the same bit $x_{i_j}$ of the
input.  Such a program is said to have {\em length $\ell$}, and {\em
width $k$} if each level has at most $k$ nodes.
\eedef

Oblivious branching programs have an elegant algebraic definition.
Recall that a {\em monoid} is a set with an associative binary
operation $\cdot$ and an identity $1$ such that $1 \cdot a = a \cdot 1
= a$ for all $a$.

\bedef 
Let $M$ be a monoid and $S \subset M$ an accepting set.  Let
$x_i$, $1 \le i \le n$ be a set of Boolean variables.  A {\em
  branching program over $M$ of length $\ell$} is a string of $\ell$
{\em instructions}; the $j$'th instruction is a triple $(i_j,a_j,b_j)
\in \{1,\ldots,n\} \times M \times M$, which we interpret as $a_j$ if
$x_{i_j}=0$ and $b_j$ if $x_{i_j}=1$.  Given an input $x$, the {\em
  yield} $Y(x)$ of the program is the product in $M$ of all its
instructions.  We say that the input $x$ is {\em accepted} if $Y(x)
\in S$, and the set of such inputs is the language recognized by the
program.  
\eedef

Such programs are often called {\em non-uniform deterministic finite
  automata} (NUDFAs).  A computation in a deterministic finite
automaton can be thought of as taking a product in its syntactic
monoid; in a NUDFA we generalize this by allowing the same variable to
be queried many times, and allowing ``true'' and ``false'' to be
mapped into a different pair of monoid elements in each query.

A common monoid is $T_k$, the set of functions from a set of $k$
objects into itself.  Then the program makes transitions among $k$
states, and we can equivalently define oblivious, width-$k$ branching
programs by choosing an initial state and a set of accepting final
states, where the $k$ states correspond to the $k$ vertices (according
to an arbitrary ordering) in each level $V_j$.

\bedef\label{oblivious}
An {\em oblivious width-$k$ branching program} is a branching program
over $T_k$, where the accepting set $S \subset T_k$ consists of those
elements of $T_k$ that map an initial state $s \in \{1,\ldots,k\}$ to
a final state $t \in A$ for some subset $A \subset \{1,\ldots,k\}$.
\eedef

\section{Bounded Width Branching Programs}

We define language classes recognized by (non-uniform) families of
bounded-width branching programs whose length increases polynomially
with $n$:


\bedef
$k$-$\BWBP$ is the class of languages recognized by polynomial-length
branching programs of width $k$, and $\BWBP = \cup_k \,k$-$\BWBP$.
\eedef

Recall that a {\em group} is a monoid where every element has an
inverse, and a group is {\em Abelian} if $ab = ba$ for all $a,b$.  A
subgroup $H \subseteq G$ is {\em normal} if the left and right cosets
coincide, $aH = Ha$ for all $a \in G$.  A group is {\em simple} if it
has no normal subgroups other than itself and $\{1\}$.

Barrington~\cite{bar85} studied branching programs over the
permutation group on $k$ objects $S_k \subset T_k$; such programs are
called {\em permutation programs}.  He showed that polynomial-length
programs over $S_5$, and therefore width-$5$ branching programs, can
recognize any language in $\NC^1$, the class of languages recognizable
by Boolean circuits of polynomial width and logarithmic depth~\cite{papa}.  
The version of Barrington's result that we will use is:
\begin{theorem}[\cite{bar85,mtlbd00}]
\label{thm:barrington}
Let $G$ be a non-Abelian simple group, and let $a \ne 1$ be any
non-identity element.  Then any language $L$ in $\NC^1$ can be
recognized by a family of polynomial-length branching programs over
$G$ such that their yield is $Y(x) = a$ if $x \in L$ and $1$
otherwise.
\end{theorem}
Since the smallest non-Abelian simple group is $A_5 \subset S_5$, the
group of even permutations of $5$ objects, and since we can choose 
a permutation $a$
that maps some initial state $s$ to some other final state $t$, width $5$
suffices.  Conversely, note that we can model a width-$k$ branching
program as a Boolean product of $\ell$ transition matrices of dimension
$k$, and a simple divide-and-conquer algorithm allows us to calculate
this product in $O(\log l)$ depth.  Thus $\BWBP \subset \NC^1$, so we
have
\[ \mbox{5-$\BWBP$} = \BWBP = \NC^1 \enspace . \]

In the stochastic and quantum cases, the state  of the
program will be described by a $k$-dimensional vector $\mu$.  The
$j$'th step of the program will query a variable $x_{i_j}$, and apply
a transition matrix $M_j(x_{i_j})$.
%
%
\bedef 
We call state (state vector) $\mu$ of branching program a
consistent state if there exists an input $x$ that induces a chain of
transitions leading to the state $\mu$ from the initial state  $\mu_0$.
Otherwise, we call the $\mu$ an inconsistent state.  
\eedef
If $x=x_1,\dots,x_n$ is the input of a program of length $\ell$, then
the final consistent state of the program will be

\[ \ket{\mu(x)}  =  \prod_{j=\ell}^1 M_j(x_{i_j}) \ket{\mu_0} \enspace . \]
(This product is in reverse order since we think of each step of the
program as a left multiplication of the state by $M_j$.)  In the
deterministic case, state $\mu$ is a Boolean vector with exactly one 1
and the $M_j$ correspond to elements of $T_k$ and so have exactly one
$1$ in each column.  For branching programs over groups this is true
of the rows as well, in which case the $M_j$ are permutation matrices.
We can generalize this by letting $\mu$ be a probability distribution,
and letting the $M_j$ be {\em stochastic} matrices, i.e.,\ matrices
with non-negative entries where each column sums to $1$, or {\em
  doubly stochastic} where both the rows and columns sum to 1.  Recall
that, by Birkhoff's Theorem~\cite{lw01}, doubly stochastic matrices
are convex combinations of permutation matrices~\cite{???}.

In the deterministic and stochastic cases for sink state vector $\mu\in
V_{\ell+1}$, we define 
\begin{equation}
\label{eq:pstoch} 
 \Pr(\mu) = \sum_{i\in \Accept} \braket{i}{\mu} 
 = \left\| \Pi_\Accept \mu \right\|_1 \enspace 
\end{equation}
and we define the probability of acceptance as $Pr(x) = Pr(\mu(x))$.
Here $\ket{i}$ is the basis vector with support on the state $i$, 
and $\Pi_\Accept$ is a projection operator on the 
{\em accepting subspace} ${\rm span}\{\ket{i}: i\in \Accept\}$.

For {\em quantum} branching programs, $\mu$ is a complex state vector
with $\| \mu \|_2 = 1$, and the $M_j$ are complex-valued unitary matrices.
 %
 %
Then we define $Pr(\mu)$ for sink state  vector $\mu\in V_{\ell+1}$ as 
\begin{equation}
\label{eq:pquant}
 \Pr(\mu)= \sum_{i\in \Accept} |\braket{i}{\mu}|^2 
 = \left\| \Pi_\Accept \mu \right\|_2^2 
\end{equation}
and the probability of acceptance as $Pr(x)=Pr(\mu(x))$
which is the probability that, that if we measure $\mu(x)$, we will
observe it in the accepting subspace.  Note that this is a
``measure-once'' model analogous to the model of quantum finite
automata in~\cite{mc97}, in which the system evolves unitarily except
for a single measurement at the end.  We could also allow multiple
measurements during the computation, by representing ${\cal A}=\{\mu :
Pr(\mu) >1/2 \}$ the state as a density matrix and making the $M_j$
superoperators; we do not consider this here.


We can define recognition in several ways for the quantum case.  We
say that a language $L$ is accepted {\em with unbounded error} if
$\Pr(x) > 1/2$ if $x \in L$ and $\Pr(x) \le 1/2$ if $x \notin L$.  We
say that a language $L$ is accepted {\em with bounded error} if there
is some $\eps > 0$ such that $\Pr(x) \ge 1/2+\eps$ if $x \in L$ and
$\Pr(x) \le 1/2-\eps$ if $x \notin L$. 
For the case $\eps=1/2$ we say $L$ is accepted {\em exactly}. That is,
$\Pr(x) = 1$ if $x \in L$ and $\Pr(x) = 0$ if $x \notin L$ as in the
deterministic case.

\section{Syntactic Variants of Stochastic and Quantum Programs}

For unbounded and bounded error stochastic and quantum branching
programs we define two subsets $\cal A$ and ${\cal R}$ of the set of
sink state vectors as follows:
For an unbounded error programs, we define

\[ {\cal A}=\{\mu\in V_{\ell+1} : Pr(\mu) >1/2 \} \qquad \mbox{and}
\qquad  {\cal
  R}=\{\mu\in V_{\ell+1} : Pr(\mu)
\le 1/2 \}; \]
for bounded error programs,we define 

\[ {\cal A}=\{\mu\in V_{\ell+1} : Pr(\mu) \ge 1/2+\eps \} \quad  \mbox{and}
\quad  {\cal
  R}=\{\mu\in V_{\ell+1} : Pr(\mu)
\le 1/2-\eps \} \]
We call $\cal A$ the accepting set of
state vectors and call $\cal R$ the rejecting set of
state vectors.

\bedef 
We call a stochastic or a quantum branching program a syntactic 
program if its accepting and rejecting set of state vectors
form a partition of the set of sink state vectors for the program. That is
$V_{\ell+1}= {\cal A}\cup {\cal R}$.  
\eedef

Observe that deterministic, unbounded error stochastic
and quantum branching programs are syntactic programs. But in the
case of bounded error branching programs it might happen that
$V_{\ell+1}\not= {\cal A}\cup {\cal R}$.  That is, it might happen
that a inconsistent final state vector $\mu\in V_{\ell+1}$ has the
property that $1/2-\eps < Pr(\mu)< 1/2+\eps$.

We denote by $B\cdot$   
the language classes recognized by standard (nonsyntactic) programs with
 bounded error  and denote by $E\cdot$  those
recognized exactly.  The notations $\SBP$ and $\QBP$ stand for
stochastic and quantum branching programs, respectively. We denote the
classes of languages recognized by width-$k$ stochastic and quantum
programs of polynomial length as $k$-$\BSBP$, $k$-$\BQBP$, and
$k$-$\EQBP$.  Note that we remove ``BW'' to avoid acronym overload. We
write $\BSBP$ for $\cup_k k$-$\BSBP$ and define $\BQBP$ and $\EQBP$
similarly.  Clearly we have
\[ \mbox{$\BWBP$} \subseteq \mbox{$\EQBP$} 
   \subseteq \mbox{$\BQBP$} 
\]
and
\[ \mbox{$\BWBP$} \subseteq \mbox{$\BSBP$} \]
but in principle $k$-$\BSBP$ could be incomparable with $k$-$\EQBP$ or
$k$-$\BQBP$.

\section{Width-$2$ Doubly Stochastic and Quantum Programs}

In this section we show that width-$2$ syntactic quantum programs
with exact acceptance contain $\nc^1$, and also that this class of
programs is stronger than width-$2$ syntactic doubly stochastic
programs.

\belem
\label{ordering}
Any width-$2$ doubly stochastic program on $n$ variables is equivalent
to one which queries each variable once and in the order
$x_1, x_2, \ldots, x_n$.
\eelem
{\em Proof.}
Any $2 \times 2$ stochastic matrix can be written as $\mat p & 1-p \\
1-p & p \rix$ for some $p \in [0,1]$. It is easy to verify that
matrices of this form commute. Hence, if we have a product of such
matrices $\prod^1_{j=n} M_j(x_{i_j})$ we can rewrite it so that we first
take the product of all the matrices that depend on $x_1$, then those
that depend on $x_2$, and so on. To finish the proof we note that
products of doubly stochastic matrices are again doubly stochastic, so
we can use a single doubly stochastic matrix for the product of all
the matrices that depend on a given $x_i$.
\Endproof

The above lemma shows we can convert any width-$2$ doubly stochastic
program into one which is read-once and with a fixed variable
ordering. i.e., a randomized ordered binary decision diagram (OBDD).
Hence in the case of width-$2$ syntactic and nonsyntactic models
of programs are equivalent. 

First we note that stochastic programs are stronger than
permutation programs for width $2$.  It is easy to see that any
program over $\Z_2$ simply yields the parity of some subset of the
$x_i$.  The $\AND_n$ function, which accepts only the input with
$x_i=1$ for all $i$, is not of this form, and so this function cannot
be recognized by a width-2 permutation program.  However, it can
easily be recognized by a  stochastic program $P$ with
bounded error which queries each variable once as follows: for $i < n$
it maps $x_i=1$ and $0$ to the
identity $\mat 1 & 0 \\ 0 & 1 \rix$ and the matrix $\mat 1/2 & 1/2 \\
1/2 & 1/2 \rix$ respectively, and for $x_n$ it maps $1$ and $0$ to
$\mat 3/4 & 0\\ 1/4 & 1 \rix$ and $\mat 3/8 & 3/8 \\ 5/8 & 5/8 \rix$
respectively. Taking the first state to be both the initial and final
state, $P$ accepts with probability $3/4$ if $x_i = 1$ for all $i$ and
$3/8$ otherwise.  Note that except for one matrix this is in fact a
doubly stochastic program; if we had treated the variable $x_n$ in the
same fashion as the other variables we would have gotten a syntactic
doubly stochastic program accepting $\AND_n$ with one-sided error.

Despite being stronger than their permutation counterparts, the next
result shows width-$2$ doubly stochastic branching programs are not that
strong. Let $\MULT^n_k$ be the Boolean function which computes the
$k$'th bit of the product of two $n$-bit integers. Define $\MULT^n$ to
be $\MULT^n_{n-1}$, i.e.,  the middle bit of the product. We will
argue that any width-$2$ stochastic program that calculates this function 
(i.e., that recognizes the set of inputs for which $\MULT^n = 1$) 
requires at least exponential width.

In \cite{ak98} Ablayev and Karpinski investigated randomized OBDDs, 
i.e., those which accept with bounded error. 
\begin{theorem}[\cite{ak98}] 
Any randomized $\OBDD$ that correctly computes $\MULT^n$ has width at least
$2^{\Omega(n/{\log n})}$.
\end{theorem}
So by Lemma~\ref{ordering} we have immediately:
\becor
$\MULT^n$ can not be computed by a width-$2$ doubly stochastic program.
\eecor

While width-$2$ doubly stochastic programs are quite weak, 
%
%
the next result shows that width-$2$ {\em quantum} programs are surprisingly 
strong.  Note that a width-$2$ quantum program has a state space
equivalent to a single qubit, such as a single spin-$1/2$ particle.

\begin{theorem}
\label{thm:width2}
$\NC^1$ is contained in syntactic $2$-$\EQBP$.
\end{theorem}

\noindent
{\em Proof.}  
Recall that the smallest non-Abelian simple group $A_5$ is isomorphic 
to the set of rotations of the icosahedron.  Therefore, the group $\SO(3)$
of rotations of $\R^3$, i.e., the $3 \times 3$ orthogonal matrices
with determinant $1$, contains a subgroup isomorphic to $A_5$.

There is a well-known 2-to-1 mapping from $\SU(2)$, the group of $2
\times 2$ unitary matrices with determinant $1$, to $\SO(3)$.
Consider a qubit $a\ket{0} + b\ket{1}$ with $|a|^2+|b|^2 = 1$; we can
make $a$ real by multiplying by an overall phase.  The {\em Bloch
sphere} representation (see e.g.~\cite{nc00}) views this state as the
point on the unit sphere with latitude $\theta$ and longitude $\phi$ ,
i.e., $(\cos\phi \cos\theta, \sin\phi \cos\theta, \sin\theta)$, where
$a \cos \theta/2$ and $b = e^{i \phi} \sin \theta/2$.

Given this representation, an element of $\SU(2)$ is equivalent to some
rotation of the unit sphere.  Recall the {\em Pauli matrices}
\[ \sigma_x = \mat 0 & 1 \\ 1 & 0 \rix, \quad
   \sigma_y = \mat 0 & i \\ -i & 0 \rix, \quad
   \sigma_z = \mat 1 & 0 \\ 0 & -1 \rix
\]
Then we can rotate an angle $\alpha$ around the $x$, $y$ or $z$ axes
with the following operators:
\begin{eqnarray*}
R_x(\alpha) = e^{i (\alpha/2) \sigma_x} & = & 
\mat \cos \alpha/2 & -i \sin \alpha/2 \\ 
  -i \sin \alpha/2 & \cos \alpha/2 \rix \\
R_y(\alpha) = e^{i (\alpha/2) \sigma_y} & = &
\mat \cos \alpha/2 & -\sin \alpha/2 \\ 
     \sin \alpha/2 & \cos \alpha/2 \rix, 
\mbox{ and} \\
R_z(\alpha) = e^{i (\alpha/2) \sigma_z} & = &
\mat e^{-i \alpha/2} & 0 \\ 
     0 & e^{i \alpha/2} \rix \enspace . 
\end{eqnarray*} 
This makes $\SU(2)$ a {\em double cover} of $\SO(3)$, where each element
of $\SO(3)$ corresponds to two elements $\pm U$ in $\SU(2)$.  (Note that
angles get halved by this mapping.)  Therefore, $\SU(2)$ has a subgroup
which is a double cover of $A_5$.  One way to generate this subgroup
is with $2\pi/5$ rotations $a$ and $b$ around two adjacent vertices of an
icosahedron.  Since two such vertices are an angle $\tan^{-1} 2$
apart, if one lies on the $z$ axis and the other lies in the
$x$-$z$ plane, we have
\begin{eqnarray*}
a & = & R_z(2\pi/5)  =  \mat e^{i \pi/5} & 0 \\ 0 & e^{-i \pi / 5} \rix \\
b & = & R_y(\tan^{-1} 2) \cdot a \cdot R_y(-\tan^{-1} 2)\\
  & = & \frac{1}{\sqrt{5}} 
  \mat e^{i \pi/5} \tau + e^{-i \pi/5} \tau^{-1} &
     - 2 i \sin \pi/5 \\
     - 2 i \sin \pi/5 &
       e^{-i \pi/5} \tau + e^{i \pi/5} \tau^{-1} \rix
\end{eqnarray*}
where $\tau = (1+\sqrt{5})/2$ is the golden ratio.  Now consider the
group element $c = a \cdot b \cdot a$; this rotates the icosahedron by
$\pi$ around the midpoint of the edge connecting these two vertices.
In $\SU(2)$, this maps each of the eigenvectors of $\sigma_y$ to the
other times an overall phase.  These eigenvectors are
\[ e_+ = \frac{\ket{0} + i \ket{1}}{\sqrt{2}},
   \quad e_- = \frac{\ket{0} - i \ket{1}}{\sqrt{2}} \]
so we have
\[ | \langle e_+ | \pm c | e_- \rangle |^2 = 1 \]
while, since they are orthogonal, 
\[ | \langle e_+ | 1 | e_- \rangle |^2 = 0 \enspace . \]

Now, Theorem~\ref{thm:barrington} tells us that for any language $L$ in
$\NC^1$ we can construct a polynomial-length program over $A_5$ that
yields the element equivalent to $c$ if the input is in the language 
and $1$ otherwise.  Using the embedding of $A_5$ in $\SO(3)$ and then 
lifting to $\SU(2)$ gives a program which yields $\pm c$ or $1$.  
If we take the initial state to be $\mu_0 = e_-$ and the 
accepting subspace to be that spanned by 
$e_+$, this program accepts $L$ exactly.
\Endproof

\section{Deterministic Simulations of Syntactic Stochastic and
  Quantum Branching Programs}

In this section we give general results on simulating syntactic
stochastic and quantum programs with deterministic ones.
Specifically, Theorem~\ref{l-ub} shows that syntactic stochastic and
quantum programs that accept with bounded error can be simulated by
deterministic programs of the same length and larger (but still
bounded) width.  Below we use this to place upper bounds on the
computational power of stochastic programs with various widths and
error thresholds.

\begin{theorem} \label{l-ub}
Let $P$ be a syntactic stochastic or quantum branching program 
of width $k$ and length $\ell$ that recognizes a language $L$
with probability $1/2+\eps$.  Then, there exists a 
deterministic branching program $P'$ of width $k'$ and length $\ell$ 
that recognizes $L$, where
\begin{equation} 
\label{eq:widthstoch}
k' \leq \left(\frac{1}{\eps}\right)^{k-1} 
\end{equation}
if $P$ is  stochastic, and 
\begin{equation}
\label{eq:widthquant}
k' \leq \left(\frac{2}{\eps}\right)^{2k} 
\end{equation}
if $P$ is quantum.
\end{theorem}

\noindent {\em Proof.}  Our proof is a simple generalization of
arguments of Rabin~\cite{rabin} and Kondacs and Watrous~\cite{kw97} 
to the non-uniform case.  
For each step of the program, we define an equivalence relation 
on state vectors, where two state vectors are equivalent if they 
lead to the same outcome (acceptance or rejection).  
Since $P$ recognizes $L$ with bounded error, inequivalent 
states must be bounded away from each other, and 
since the state space is compact the number of
equivalence classes is finite.  These equivalence classes then
become the states of our deterministic program $P'$.  


First, we construct a much larger deterministic branching program
$P''$ whose states at each level $j$ consist of the state vectors
(consistent and nonconsitent) $\mu$ that $P$ can reach after $j$ steps
(computational or noncomputational).
$V''_1$ consists of 
the initial state $\mu_0$, and for all $1 \le j \le \ell$ each 
$\mu \in V''_j$ has two outgoing edges to
$M_j(0) \mu, M_j(1) \mu \in V''_{j+1}$.
The syntactic property provides partition the final states $\mu
\in V''_{\ell+1}$ into accepting and rejecting subsets $\cal A$ and
$\cal R$ according to equations~\eqref{eq:pstoch}
and~\eqref{eq:pquant}.

In the stochastic case, $\mu \in {\cal A}$ (resp.\ $\mu \in {\cal R}$) 
if $\| \Pi_\Accept \mu \|_1 \ge 1/2+\eps$
(resp.\ $\| \Pi_\Accept \mu \|_1 \le 1/2-\eps$), and 
similarly in the quantum case except that $\|\cdot\|_1$ 
is replaced with $\|\cdot\|_2^2$.
Clearly if ${\cal A}$ is the accepting subset of $V''_{\ell+1}$, 
then $P''$ recognizes $L$ deterministically.

Now we inductively define an equivalence relation $\equiv_j$
on $V''_j$ for each level $j$.  First, we let 
${\cal A}$ and ${\cal R}$ be the equivalence classes of $\equiv_{\ell+1}$
(note that  $V''_{\ell+1} = {\cal A} \,\cup\, {\cal R}$
since $P$ recognizes $L$ with bounded error).
Then, for each $1 \le j \le \ell$, define 
\[ \mu \equiv_j \mu' \;\Leftrightarrow\; M_j(0) \mu \equiv_{j+1} M_j(0) \mu'
 \,\mbox{ and }\, M_j(1) \mu \equiv_{j+1} M_j(1) \mu' 
\enspace . \]
We now define $P'$ as the branching program whose states 
$V'_j$ for each level $j$ are the equivalence classes of $\equiv_j$ 
and whose accepting subset is the singleton $\{ {\cal A} \}$.
Clearly $P'$ is well-defined and recognizes $L$ deterministically; 
it just remains to show that the number of equivalence classes 
for each $j$ is bounded.

First we show that two inequivalent state vectors in $V''_j$ must
be far apart, using the following standard argument~\cite{rabin,kw97}.
\belem\label{theta} Suppose $\mu, \mu' \in V''_j$
and $\mu \not \equiv_j \mu'$.  Then 
\[ \|\mu-\mu'\|_1 \geq 4 \eps \]
if $P$ is stochastic, and
\[ \|\mu-\mu'\|_2 \geq 2 \eps \]
if $P$ is quantum.
\eelem

{\noindent \em Proof.}  Since stochastic and unitary matrices both
preserve or decrease the appropriate norm, it suffices to show this
for the last step.  Therefore, suppose that $j=\ell+1$, $\mu \in {\cal
  A}$ and $\mu' \in {\cal R}$.  We can decompose both vectors into
their components inside the accepting subspace and transverse to it,
writing $\mu = \mu_A + \mu_R$ where $\mu_A = \Pi_\Accept \mu$ and
$\mu_R = (1-\Pi_\Accept) \mu$ and similarly for $\mu'$.  In the
stochastic case, $\|\mu_A\|_1 \ge 1/2+\eps$ and $\|\mu'_A\|_1 \le
1/2-\eps$, and so

\begin{eqnarray*}
\|\mu- \mu'\|_1 & = & \| \mu_A - \mu'_A \|_1 + \| \mu_R - \mu'_R \|_1 \\
& \geq & 2 \left[ \left(1/2+\eps\right) - \left(1/2-\eps\right) \right] \\
& = & 4 \eps \enspace . 
\end{eqnarray*}
In the quantum case, $\|\mu_A\|_2 \ge \sqrt{1/2+\eps}$ and 
$\|\mu'_A\|_2 \le \sqrt{1/2-\eps}$, so
\begin{eqnarray*}
\|\mu-\mu'\|_2^2 & = & \|\mu_A-\mu'_A\|_2^2 + \|\mu_R-\mu'_R\|_2^2 \\
& \ge & 2 \left[ \sqrt{1/2+\eps} - \sqrt{1/2-\eps} \right]^2 \\
& = & 2 \left( 1 - \sqrt{1-4\eps^2} \right) \\
& \ge & 4\eps^2
\end{eqnarray*}
so $\|\mu-\mu'\|_2 \ge 2\eps$.
\Endproof

It follows that the width $k'$ of $P'$ is at most the largest number
of balls of radius $2 \eps$ or $\eps$ (in the stochastic and quantum
case respectively) one can fit inside the state space. 
In the stochastic case, the state space is a $(k-1)$-dimensional
simplex.  Its $L_1$-diameter is $2$, so each ball of radius $2\eps$
covers a fraction at least $(1/\eps)^{k-1}$ of its volume,
yielding~\eqref{eq:widthstoch}.  This bound is crude in that it
assumes that the center of each ball is at a corner of the simplex;
balls whose center are in the interior of the simplex cover up to
$2^{k-1}$ times as much volume.  In particular, if $k=2$ then $k' \le
1+1/(2 \eps)$.

In the quantum case, the state space is isomorphic to the surface of
the $2k$-dimensional sphere of radius $1$.  The crude bound 
of~\eqref{eq:widthquant} comes from noticing that this sphere, 
and the balls of radius $\eps$ whose centers lie on its surface, 
are all contained in a $2k$-dimensional ball of radius $2$. 
\Endproof

Theorem~\ref{l-ub} shows that bounded-error syntactic stochastic and
quantum programs of constant width can be simulated by deterministic
programs of constant (albeit exponentially larger) width, and are
therefore contained in $\NC^1$.  Conversely, we showed in
Theorem~\ref{thm:width2} that $\NC^1$ is contained in width-$2$
syntactic quantum programs.  Therefore, the following classes all
coincide with $\NC^1$:
\becor
For syntactic programs,
\[ \mbox{$2$-$\EQBP$} = \mbox{$2$-$\BQBP$} = \EQBP = \BQBP = \BSBP = \BWBP 
= \NC^1 \enspace .
\]
\eecor
Of all the program classes discussed in this paper, the only ones {\em
  not} included in this collapse are stochastic programs of width less
than $5$.  Theorem 4 allows us to place upper bounds on their
computational abilities if their error margins are sufficiently large.
For instance, since Yao~\cite{yao83} showed that width-2 deterministic
programs require superpolynomial length to compute the majority
function, we have 
\becor For the syntactic case, width-$2$ stochastic branching programs of
polynomial length cannot recognize the majority function with
probability $1/2+\eps$ if $\eps > 1/4$.  \eecor

Similarly, recall that $\ACC = \cup_p \ACC[p]$ where 
$\ACC[p]$ is the class of languages 
recognizable by constant-depth circuits with AND, OR, and 
mod-$p$ counting gates of arbitrary fan-in.  It is known that 
$\ACC[p] \subsetneq \NC^1$ for prime $p$~\cite{razborov,smolensky}, 
and strongly believed, but not known, that $\ACC \subsetneq \NC^1$.
Since its is known~\cite{bt88} that deterministic programs of width 
less than $5$ recognize languages in $\ACC[6]$, we have
\becor
Suppose $L$ is recognized with probability $1/2+\eps$ by a
width-$k$ stochastic syntactic branching program of polynomial length.  
If $k=2$ and $\eps > 1/8$, or $k=3$ and $\eps > 1/3$, 
or $k=4$ and $\eps > 3/8$, then $L \in \ACC$.
\eecor

\noindent
{\em Proof.}  For each $k$ we consider the problem of how small 
$\eps$ has to be to fit $5$ points into the $(k-1)$-dimensional simplex
with an $L_1$ distance $4\eps$ between them.  While these values of $\eps$ are
smaller that those given by~\eqref{eq:widthstoch}, they follow easily from 
assuming without loss of generality that $k$ of the points lie on the 
simplex's corners.
\Endproof

However, we conjecture that stochasticity doesn't greatly increase the
power of bounded-width branching programs, in the following sense:
\begin{conjecture}
If $L$ is recognized with bounded error by a stochastic branching program
of width less than $5$, then $L \in \ACC$.
\end{conjecture}

\bigskip {\bf Acknowledgments.}  We grateful to Sasha Razborov for
numeric suggestions on the results presentations. We thank Denis
Th\'erien for helpful discussions on $\ACC$ and Sasha Shen for
productive discussions on syntactic model.

\bibliographystyle{plain}

\end{document}